\documentclass[aip,amsmath,amssymb,reprint,]{revtex4-1}

\usepackage{graphicx}
\usepackage{dcolumn}
\usepackage{bm}
\usepackage[utf8]{inputenc}
\usepackage[T1]{fontenc}
\usepackage{mathptmx}
\usepackage{etoolbox}
\usepackage{longtable}
\usepackage{xcolor}
\usepackage{ulem}

\makeatletter
\def\@email#1#2{%
 \endgroup
 \patchcmd{\titleblock@produce}
  {\frontmatter@RRAPformat}
  {\frontmatter@RRAPformat{\produce@RRAP{*#1\href{mailto:#2}{#2}}}\frontmatter@RRAPformat}
  {}{}
}%
\makeatother
\begin{document}

\preprint{AIP/123-QED}

\title[Twitch chats exhibit multifractality]{Some Twitch chats exhibit multifractal characteristics resembling stream-of-consciousness narrative}

\author{Marcin Ból}
\affiliation{Faculty of Computer Science and Mathematics, PK Krakow University of Technology, ul. Warszawska 24, 31-155 Kraków, Poland}

\author{Stanisław Drożdż}
\email{stanislaw.drozdz@ifj.edu.pl}
\affiliation{Faculty of Computer Science and Mathematics, PK Krakow University of Technology, ul. Warszawska 24, 31-155 Kraków, Poland}
\affiliation{Complex Systems Theory Department, Institute of Nuclear Physics, Polish Academy of Sciences, ul. Radzikowskiego 152, 31-342 Kraków, Poland}

\author{Robert Kluszczyński}
\affiliation{Complex Systems Theory Department, Institute of Nuclear Physics, Polish Academy of Sciences, ul. Radzikowskiego 152, 31-342 Kraków, Poland}
\affiliation{Faculty of Mathematics and Computer Science, Jagiellonian University, ul. Łojasiewicza 6, 30-348 Kraków, Poland}

\author{Jarosław Kwapień}
\affiliation{Complex Systems Theory Department, Institute of Nuclear Physics, Polish Academy of Sciences, ul. Radzikowskiego 152, 31-342 Kraków, Poland}

\author{Marcin Wątorek}
\affiliation{Faculty of Computer Science and Mathematics, PK Krakow University of Technology, ul. Warszawska 24, 31-155 Kraków, Poland}

\date{\today}

\begin{abstract}
Live-streaming platforms generate high-frequency records of collective human activity in which many loosely coordinated users react in real time to shared external stimuli and to one another. We investigate whether the temporal organization of Twitch chat messages exhibits multifractal signatures characteristic of complex dynamical systems. Selected high-activity chats were converted into time series by measuring the length of each consecutive message in words, with platform emotes treated as individual tokens. The resulting signals display heterogeneous fluctuations, bursty organization, and heavy-tailed distributions that, in several cases, are better captured by q-Weibull forms than by conventional light-tailed models. Autocorrelation analysis reveals persistent temporal dependence over broad ranges of message lags, indicating that the variability is not produced by independent message generation alone. Multifractal Detrended Fluctuation Analysis shows that some chats possess well-developed multifractal spectra and generalized Hurst exponents strongly dependent on moment order, whereas others exhibit weak scaling, approximately monofractal behavior, or no well-defined scaling regime. Surrogate tests indicate that temporal correlations are essential for the observed multifractality, while heavy tails enhance its apparent strength. Under particular dynamical conditions, Twitch chat may therefore form an emergent collective signal whose fragmented, reactive, associative, and multiscale organization is structurally reminiscent of stream-of-consciousness narrative. This expression is used strictly as a structural and dynamical analogy, not as a claim of a unified collective mind or of semantic, cognitive, or phenomenological equivalence with literary stream-of-consciousness writing. The results position live-stream chat dynamics within the broader physics of complex, intermittent, human-generated signals.
\end{abstract}

\maketitle

\begin{quotation}
The rapid expansion of digitally mediated communication has given rise to new forms of collective human activity that unfold at high temporal resolution and involve large numbers of interacting individuals. Live-streaming platforms, in particular, host dense streams of user-generated messages that evolve in real time in response to shared external stimuli and endogenous social dynamics. These environments provide a natural laboratory for the study of complex systems, where heterogeneous agents interact under minimal coordination yet generate rich, strongly fluctuating temporal patterns. Capturing and characterizing such variability constitutes a challenge for conventional statistical approaches, which often rely on assumptions of homogeneity or characteristic scales.

Within the broader study of nonlinear dynamics, multifractality has emerged as a universal formal tool for describing systems in which fluctuations are organized across multiple scales. Originally developed in the context of physical processes such as turbulence, multifractal analysis has since been applied to a wide range of natural and human-generated signals, offering a compact way to quantify heterogeneity, intermittency, and scale-dependent structure. As digital social interactions increasingly resemble high-frequency, nonstationary signals produced by many interacting components, multifractal frameworks offer a natural bridge between modern online communication and established theories of complex, scale-invariant phenomena.

\end{quotation}

\section{Introduction}
Digital communication platforms have become central arenas for human social interaction, generating unprecedented volumes of temporally resolved, user-driven data~\cite{LazerD-2009,RahwanI-2019}. From microblogging services to collaborative knowledge platforms, these environments capture spontaneous collective behavior at scales and speeds previously inaccessible to social science~\cite{vanDijck-2018,Abu-SalihB-2021}. More generally, the application of concepts and methods from statistical physics to collective human behavior has developed into the broad field of social and complex-systems physics \cite{JusupM-2022}. Among the related platforms, live-streaming services represent a particularly dynamic form of mediated interaction, where large audiences engage in real time with content creators and with one another through rapidly evolving textual exchanges known as stream chats. Despite their growing cultural and economic importance, the structural and dynamical properties of these interactions remain insufficiently understood.

Twitch, one of the largest live-streaming platforms worldwide, hosts millions of concurrent viewers who participate in fast-paced chat conversations synchronized with live audiovisual content. Unlike traditional online forums or social media timelines~\cite{JohnsonMR-2019}, Twitch chat is characterized by extreme temporal density, high participant turnover, and a strong coupling between external stimuli (the stream content) and endogenous social responses~\cite{Hilvert-BruceZ-2018}. These chats often include a mixture of natural language, emotes, memes, and platform-specific symbols, forming a hybrid communication system that blends linguistic expression with performative and affective signaling. As such, Twitch chat constitutes a novel social activity that reflects the diversity, spontaneity, and complexity of contemporary human interactions in digital environments~\cite{JodenStrandell2022,RecktenwaldD-2017}.

Chat conversations often consist of irregular bursts of messages triggered by gameplay events, streamer interactions, memes, or emotional reactions, producing fluctuations that span multiple temporal scales. Such patterns can emerge from heterogeneous participation—ranging from highly active users to occasional contributors—and from cascades of interaction where messages rapidly trigger further responses. Together, these mechanisms suggest that Twitch chats may function as a self-organizing social system, where collective attention and interaction spontaneously structure the flow of communication over time. From a more general perspective, collective human behavior may involve higher-order interactions among groups of individuals that cannot be reduced to pairwise relationships and may generate emergent dynamical patterns at the group level \cite{BattistonF-2025}.
In quantitative terms such a complex, multiscale dynamics may result in appearance of multifractality as in many systems with interactions across a wide range of temporal or spatial scales~\cite{KwapienJ-2012a}. In the present context, it seems particularly thought-provoking that the sentence length variability in texts belonging to the literary style known as stream of consciousness~\cite{HumphreyR-1958} exhibits multiscale organization resulting even in multifractality~\cite{DrozdzS-2016a,StaniszT-2024a,DolinaM-2025}. This style involves presenting the direct flow of a character's thoughts, feelings, memories, and associations as they arise in their mind. The text attempts to imitate the natural flow of human thought. A prime example of this type of text is {\it Finnegans Wake} by James Joyce~\cite{DrozdzS-2016a,StaniszT-2024,BartnickiK-2025}.

This study hypothesizes that multitude of high-velocity interactions in Twitch chat can exhibit structural features analogous to the literary technique of stream of consciousness~\cite{HumphreyR-1958}. Specifically, the rapid sequence of short, reactive, and loosely connected messages produced by multiple users may generate a form of collective stream-of-consciousness discourse, characterized by fragmentation, associative topic shifts, and immediate emotional expression. Such chat dynamics may therefore function as an emergent, crowd-generated narrative layer accompanying live streams~\cite{YenR-2023}.

From a quantitative perspective, Twitch chat activity can be viewed as a complex time series generated by many interacting agents operating under heterogeneous motivations, attention spans, and communicative norms. Prior research on online communication has largely focused on aggregate statistics such as message volume, sentiment, or network structure~\cite{LazerD-2009,GonzalesLA-2011,BoydD-2012}. These approaches have yielded valuable insights but they often overlook fine-scale temporal fluctuations and scaling properties that may encode deeper organizational principles of collective behavior. In particular, the question of whether chat dynamics exhibit signatures of long-range correlations or hierarchical temporal organization remains largely unexplored.

Multifractal analysis provides a unified  framework for investigating such properties. Originally developed in the context of turbulence and statistical physics, multifractal methods have since been successfully applied to a wide range of natural and social phenomena~\cite{StanleyHE-1988}, including financial markets~\cite{JiangZQ-2019a,WatorekM-2021a}, global grain spot market~\cite{GaoX-L-2022,WangL-2023}, physiological signals~\cite{IvanovPC-1999,Watorek2024}, climate records~\cite{AshkenazyY-2003}, human mobility patterns~\cite{MurcioR-2015} and even in the literary texts~\cite{DrozdzS-2016a}. A multifractal signal is characterized not by a single scaling exponent, but by a spectrum of exponents~\cite{HalseyTC-1986a} that capture variability across multiple temporal scales. The presence of multifractality is often interpreted as evidence of heterogeneous dynamics, intermittent bursts, and the coexistence of multiple interaction regimes within a single process~\cite{JiangZQ-2019a}.

In social systems, multifractality has been associated with diversity in human behavior, adaptive responses to external stimuli, and the interplay between individual actions and collective organization. For example, multifractal patterns in communication or activity time series may reflect differences in user engagement, attention cycles, or coordination mechanisms that operate across short and long timescales. 

In social and behavioral systems, multifractal scaling has been interpreted as a signature of heterogeneous variability and interactions among processes operating across multiple temporal scales \cite{IhlenVereijken2013,KeltyStephen2023}. Related fractal analyses of culturomic time series have shown that long-term word-use trajectories encode nontrivial temporal organization, with social phenomena frequently displaying nonstationary or intermittent scaling behavior \cite{GaoJ-2012}. Multifractality and long-range temporal dependence have been reported in inter-event times associated with human decisions, telephone-call activity, social-media activity, and collective online knowledge production \cite{Perello2008,Gui2021,Oh2022,Tadic2017}. At the community level, persistent temporal correlations may emerge from the collective organization and coordination of many individual activities rather than from individual inter-event-time distributions alone \cite{Rybski2012}. Such scaling patterns may therefore reflect heterogeneous participation, bursty user engagement, attention cycles, adaptive responses to external events, and coordination mechanisms acting across short and long time scales~\cite{Barabasi2005,Karsai2012,Garas2012}. Applying this framework to Twitch chat offers an opportunity to quantitatively assess whether the apparent chaos of real-time online conversations conceals structured, scale-dependent regularities.

The present study investigates the temporal dynamics of Twitch chat activity through the formalism of multifractal analysis. By treating chat message streams as time-ordered sequences and examining their scaling properties, we aim to determine whether these interactions exhibit multifractal characteristics and, if so, what this reveals about the underlying social processes. Importantly, our focus is not on individual users or semantic content, but on the collective temporal structure emerging from many simultaneous participants. This perspective aligns with the view of Twitch chat as a collective behavioral signal shaped by both platform constraints and human interaction patterns.

Identifying multifractal properties in Twitch chat would have several implications. First, it would suggest that live-stream chat dynamics cannot be adequately described by simple random or monofractal processes, highlighting the presence of rich temporal heterogeneity. Second, it would support the interpretation of Twitch chat as a complex adaptive system, where interactions across scales—from rapid bursts of reactions to slower engagement cycles—coexist and influence one another. Third, it would position live-stream chats alongside other human-generated signals known to exhibit multifractality, reinforcing the idea that digital communication platforms capture fundamental aspects of human social behavior~\cite{OksanenA-2024}.

Beyond methodological contributions, this work carries potential of addressing broader questions about how humans interact in large-scale, technologically mediated settings. Twitch chat exemplifies a form of collective expression that is ephemeral, performative, and deeply social, yet largely unstructured by explicit rules or hierarchies. The emergence of multifractal patterns in such an environment would underscore the capacity of human groups to self-organize across multiple temporal scales, even in settings dominated by noise, rapid turnover, and minimal coordination.

In the present contribution we document an empirical analysis of several exemplary Twitch chat time series drawn from a representative set of live streams. We apply established multifractal techniques to characterize their scaling behavior and in some cases identify a fully developed multifractality. However, such characteristics are not common. In many cases, uniform scaling is at most monofractal or absent altogether. This study thus contributes to the growing interdisciplinary literature at the intersection of computational social science, complex systems, and digital media studies. In particular it sheds light on how contemporary online platforms reflect the diversity and complexity of human interactions, often revealing structured dynamics within what might otherwise appear as unstructured digital chatter.

\section{Twitch data specification}

Twitch is a major interactive livestreaming platform focused primarily on video-game broadcasting, e-sports events, and creative or conversational content. The service originated as a gaming-oriented spin-off of the broader lifecasting site Justin.tv, founded in 2011 by Justin Kan and Emmett Shear. In 2014, the platform was acquired by Amazon and reorganized as Twitch Interactive, further accelerating its infrastructural development and global reach~\cite{SpilkerHS-2020,ChalabyJK-2024}.


\begin{figure*}[!htb]
    \centering
    \includegraphics[width=1\textwidth]{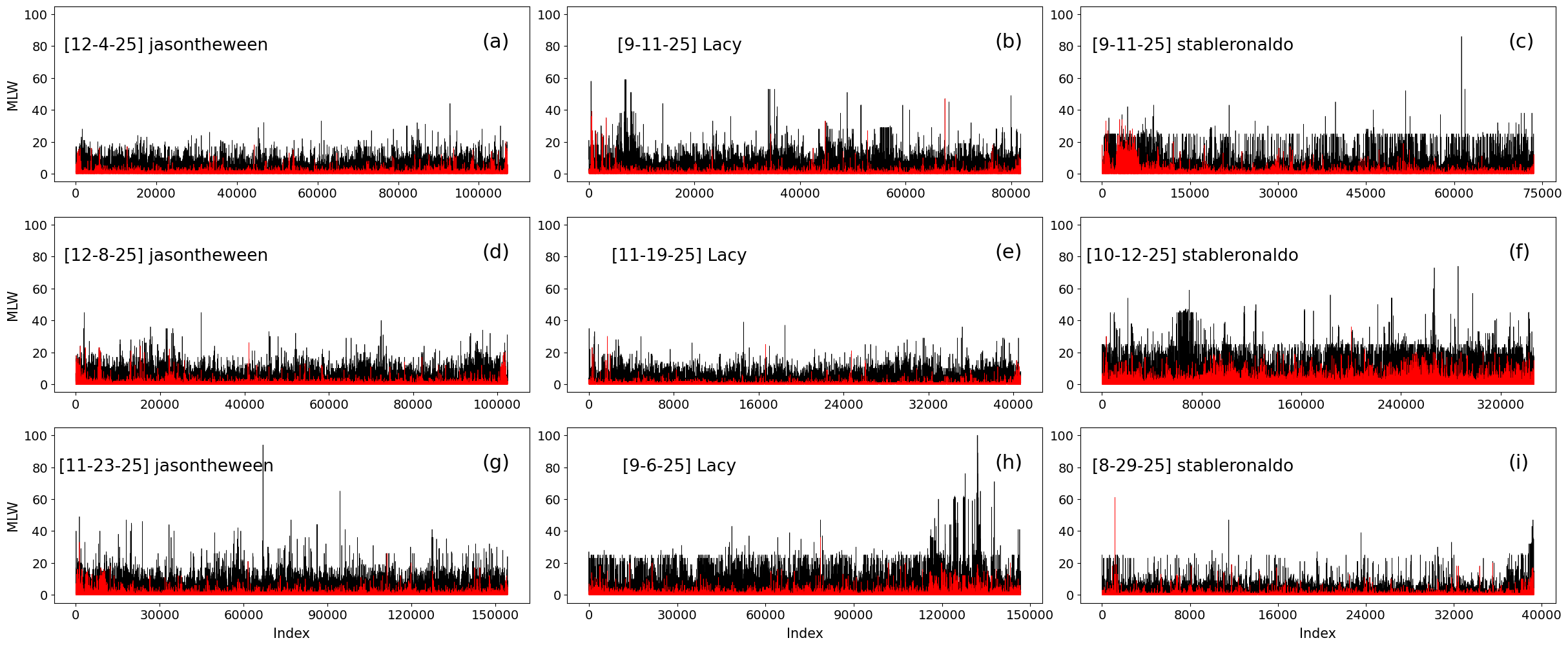}
    \caption{Lengths of the consecutive messages measured in terms of the number of words and emotes (MLW - Message Length in Words) are represented by the black vertical lines and the corresponding contribution of the number of emotes in every message are reflected by the red vertical lines. For each of the three selected top streamers, i.e., jasontheween, Lacy and stableronaldo, three different sessions, so nine in total, are illustrated in an analogous way. The dates (month-day-year) of their recording are explicitly listed in the corresponding panels.}
    \label{fig:wideplot}
\end{figure*}

Since its launch, Twitch has experienced substantial growth. Early adoption was driven by the rise of e-sports and the increasing cultural significance of real-time spectator gaming. By the mid-2010s, Twitch had become the dominant platform in the game-streaming sector, regularly outperforming competitors in total hours watched and average concurrent viewership. The COVID-19 pandemic in 2020 produced an additional surge in usage, as lockdowns increased both content production and audience engagement. Although growth has stabilized in recent years, Twitch remains one of the most popular livestreaming platforms worldwide, consistently attracting millions of daily active users and serving as a central hub for digital gaming communities, online subcultures, and emerging creator-economy ecosystems~\cite{JohnsonMR-2019}.

Messages in Twitch chat are typically short, reactive units composed of a mixture of natural language tokens, platform-specific emotes, and repeated symbolic characters. Empirical observations indicate that emotes constitute a substantial fraction of the token distribution (often exceeding 30$\%$ in high-traffic channels; see Appendix), while lexical content is sparse and highly informal, with frequent use of abbreviations and minimal syntactic structure. 

At the collective level, chat dynamics are dominated by rapid message turnover, high redundancy, and bursty repetition, producing temporally clustered patterns of identical or near-identical messages (e.g., emote chains or repeated punctuation). This structure reflects a form of synchronized crowd signaling, in which information content is secondary to the amplification of shared reactions, and leads to strong short-timescale correlations in both token usage and message frequency across users. These properties can be further amplified in a cascade to longer time scale by the high-volume, fast-paced nature of the medium and the continuously evolving vocabulary of emotes and neologisms.

In the present study TwitchDownloader~\cite{TwitchDownloader} to download Twitch chats related to top 20 most subscribed Twitch streamers of all time~\cite{TwitchTopSubcribedStreamers} was used. The chats were converted to time series by counting the number of words in every message for every collected chat. By word any sequence of characters separated by spaces is meant. This metric is particularly useful in the present analysis, since every emote present in chat is treated as a single word. It should be mentioned that emotes on twitch are typed in chat in a form of a word like "PogChamp" or "PoroSad", and then such special phrase is automatically converted to an image or a short GIF. Such a metric choice is also robust to links occasionally pasted in chats, since they are also treated as a single word. 


\begin{figure*}[!htb]
    \centering
    \includegraphics[width=1\textwidth]{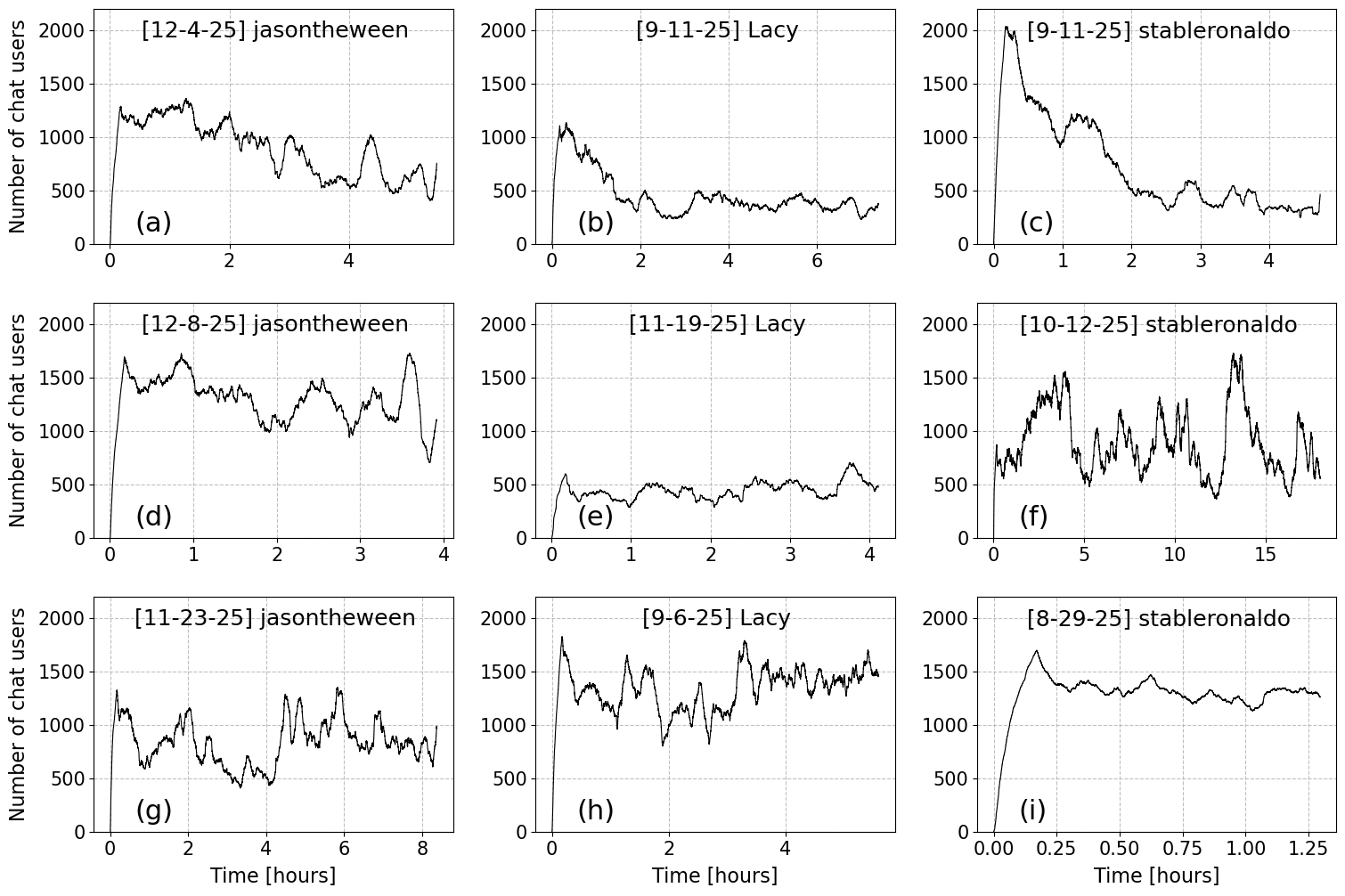}
    \caption{Time dependence of the number of unique chat participants in rolling 10 min window for each of the nine streams of Figure~\ref{fig:wideplot}.}
    \label{fig:num_of_users}
\end{figure*}

Systematic searches allowed us to identify several streamers among a dozen or so top, whose chats belong to those that reveal a large diversity of temporal correlations including long-range linear as well as nonlinear ones. The three representative streamers whose chats meet such a correlation diversity criterion were selected for the analysis presented here. These are: jasontheween, Lacy and stableronaldo. For each of them, three chats revealing the most diverse forms of nonlinear correlations appearing in the entire collection were selected, which leads to a total of nine chats considered here. The specification of their corresponding quantitative characteristics is detailed in Appendix. The anonymized dataset can be downloaded from the open repository~\cite{DataSet}.

The variability of the message of these chats flows are shown in Fig.~\ref{fig:wideplot}. The black vertical lines reflect the lengths MLW (Message Length in Words) of subsequent messages measured in word count. The apparent length of such chats is seen typically between tens of thousands and one hundred and tens of thousands of messages. The superimposed red lines illustrate the relative contribution of emotes in these messages. As can be clearly seen, the patterns defined by the variation in the lengths of successive messages have a very complex organization, both in terms of the quantitative relationships between these lengths and in terms of their distribution over time. These patterns are clearly richer in terms of organization than, for example, white noise, for which the waveforms are more uniform. Here, the tendency to cluster this variability and burstiness effects like in financial~\cite{KwapienJ-2012a,WatorekM-2021a} or some literary~\cite{DrozdzS-2016a} time series revealing multiscaling are quite clearly visible. In all three of the stableronaldo cases shown, and the one in Lacy's (the lowest in this figure), one sees the superimposed series of more evenly distributed messages with two threshold lengths of 23 and 25. Almost all of these messages appear to be automatically generated by bots, most often in response to a subscription payment. Their share of the total session length is not very large; typically it does not exceed $0.5\%$.

Complementary information revealing the variability over time of the number of active participants for the same chat composition is displayed in Figure~\ref{fig:num_of_users}. In cases (a), (b), and (c), after a very rapid increase to levels over 1,000 within the first dozen or so minutes, the number drops slightly and saturates at around 500 active participants. For the remaining (d)-(i) cases, the number of active chat participants does not follow such phases and typically assume higher values  except for case (e), where the initial increase is weaker and does not significantly exceed 500.

\section{Distribution of fluctuations}

\begin{figure*}[!htb]
    \centering
    \includegraphics[width=1\textwidth]{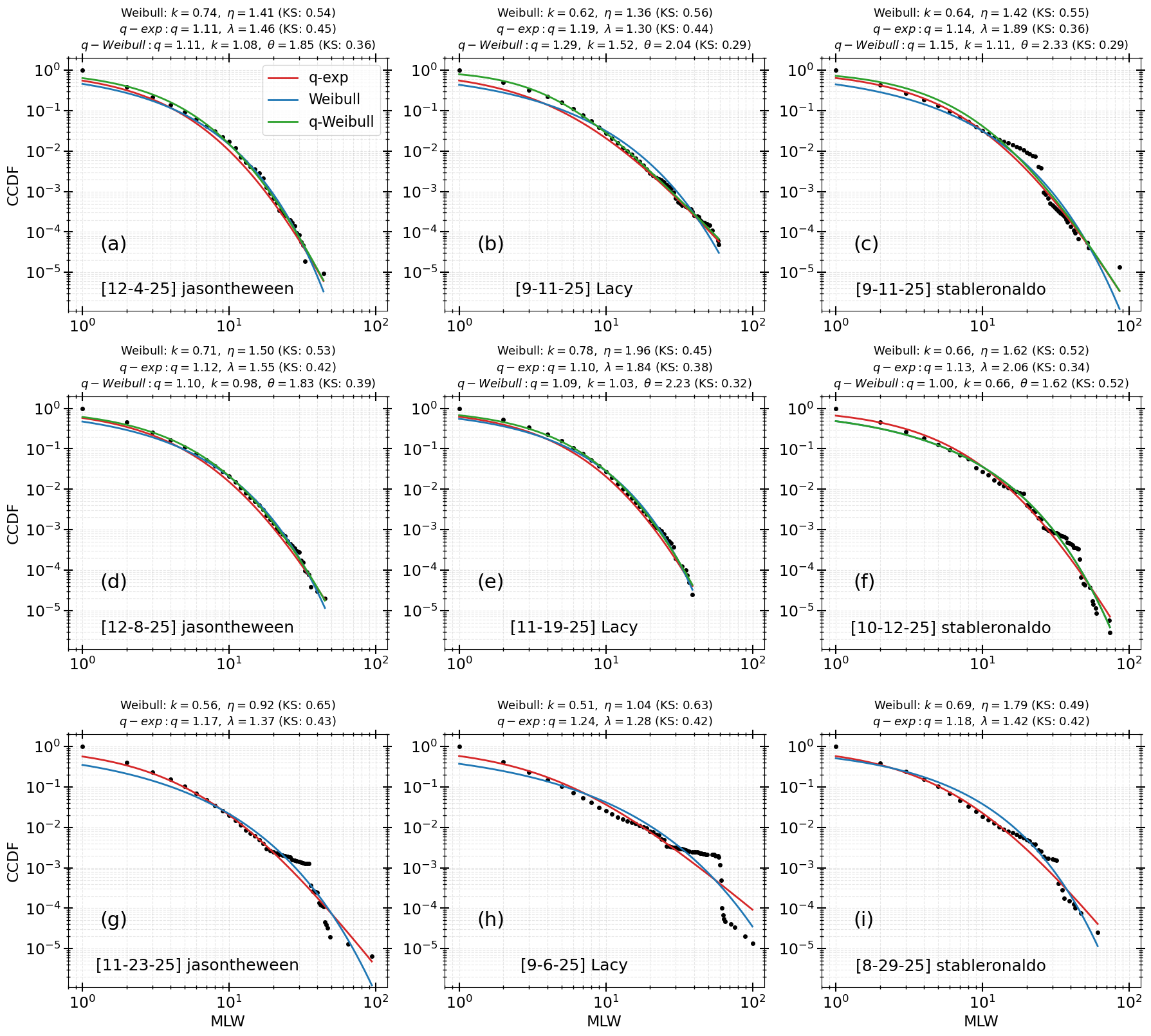}
    \caption{Complementary cumulative distributions (CCDFs) of the empirical fluctuations (dots) of message length expressed in terms of words and emotes for all nine chats studied in the same settings as in~Fig.~\ref{fig:wideplot}. Their best fits in terms of the $q$-exponential (red lines), Weibull (blue lines) and $q$-Weibull (green lines) functions are also shown.}
    \label{fig:CCDFs}
\end{figure*}

When dealing with fluctuations as in Fig.~\ref{fig:wideplot}, it is natural to ask what the distribution of such fluctuations is and what the functional form is that constitutes the best fit to these distributions. The empirical distributions of MLW, expressed in the complementary cumulative representation $\bar {F}(x)$, for all nine cases shown in Fig.~\ref{fig:wideplot} in the same setting are shown in Fig.~\ref{fig:CCDFs}. The log-log scale used quite clearly indicates that these distributions are not power-law scaling but that they are significantly heavier-tailed than the Gaussian ones, however. 


There are two theoretical distributions that may be considered in the context of fitting the present data. One is the Weibull distribution~\cite{WeibullW-1951} 
\begin{equation}
f_W(x) = \frac{k}{\eta} {\left(\frac{x}{\eta}\right)^{k-1}} \exp\! \left [{-\left(\frac{x}{\eta}\right)^k} \right ], 
\end{equation}
which has been shown~\cite{StaniszT-2023} to be very efficient in fitting the inter-punctuation intervals in written texts. Here $k>0$ denotes the shape parameter and $\eta > 0$ the scale parameter. In a complementary cumulative form (survival function) it reads  
\begin{equation}
\bar{F}_W(x) = P(X>x) = \exp\! \left [{-\left(\frac{x}{\eta}\right)^k} \right ].
\end{equation}
Another is the $q$-exponential distribution which derives from the non-extensive statistical mechanics of Tsallis~\cite{TsallisC-1988} and makes use of the generalized exponential function 
\begin{equation}
\exp_q(x) = \left[1+(1-q)x\right]^{\frac{1}{1-q}}.
\end{equation}
This distribution has thus a form
\begin{equation}
f_q(x) = \frac{2-q}{\lambda} \exp_q\!\left(-\frac{x}{\lambda}\right)
= \frac{2-q}{\lambda} \left[1-(1-q)\frac{x}{\lambda} \right]^{\frac{1}{1-q}}
\end{equation}
and in complementary cumulative representation it reads
\begin{equation}
\bar{F}_q(x) = P(X>x) = \left[\exp_q\!\left(-\frac{x}{\lambda}\right) \right]^{2-q}
= \left[1-(1-q)\frac{x}{\lambda} \right]^{\frac{2-q}{1-q}}.
\end{equation}
Its asymptotic power-law form for heavy tailed case $1 < q < 2$ is
\begin{equation}
\bar{F}_q(x) \sim x^{-\frac{2-q}{q-1}}, \qquad x\to\infty,
\end{equation}
and thus allows thicker tails than the Weibull distribution which in such a limit decays exponentially. 

When applied to the empirical data (Fig.~\ref{fig:wideplot}) distributions shown in Fig.~\ref{fig:CCDFs} (dots), both of these theoretical distributions provide satisfactory fits in the mean. It is quite apparent, however, that while the Weibull distribution (green lines) handles the mid-range fluctuation values well it tends to underestimate the probability of fluctuations at both the small and large extremes of variability. In these areas, the $q$-exponential (blue lines) seems more appropriate. From this perspective, using a distribution that combines the properties of both of these probability distributions seems to be the right step. Such a distribution, called the q-Weibull, has indeed been successfully proposed~\cite{PicoliJrS-2003} in the scientific literature.

The corresponding derivation is based on the observation that both these distributions can be written~\cite{PicoliJrS-2003,JoseKK-2009,AssisEM-2013} in the unified survival form
\begin{equation} 
\bar F(x)=\exp[-H(x)],
\end{equation}
where \(H(x)\) is the cumulative hazard function. For the Weibull distribution 
\begin{equation}
H_{\mathrm W}(x) = \left(\frac{x}{\eta}\right)^k ,
\end{equation}
whereas for the q-exponential distribution 
\begin{equation}
H_q(x) = \frac{1}{q-1}\ln\!\left[1+(q-1)\frac{x}{\lambda}\right].
\end{equation}

Thus the $q$-Weibull is obtained by replacing the Weibull cumulative hazard with its $q$-deformed version
\begin{equation}
H_{qW}(x) = \frac{2-q}{q-1}\ln\!\left[1+(q-1) H_W(x) \right].
\end{equation}
The corresponding probability functional is
\begin{equation}
f_{qW}(x)= (2-q)\frac{k}{\theta} \left(\frac{x}{\theta}\right)^{k-1} \left[1-(1-q)\left(\frac{x}{\theta}\right)^k
\right]^{\frac{1}{1-q}}, \qquad x\ge 0,
\end{equation}
or using the $q$-exponential notation
\begin{equation}
f_{qW}(x)= (2-q)\frac{k}{\theta} \left(\frac{x}{\theta}\right)^{k-1} \exp_q\!\left[-\left(\frac{x}{\theta}\right)^k\right].
\end{equation}
Here, $\theta$ denotes the scale factor which in principle can assume different values than the two previous ones. 
The resulting CCDF reads
\begin{equation}
\bar F_{qW}(x) = P(X>x) = \left[1-(1-q)\left(\frac{x}{\theta}\right)^k \right]^{\frac{2-q}{1-q}},
\end{equation}
or equivalently
\begin{equation}
\bar F_{qW}(x) = P(X>x) = \left[\exp_q\!\left(-\left(\frac{x}{\theta}\right)^k \right) \right]^{\,2-q}.
\end{equation}
From this insightful perspective the q-Weibull is a logarithmically deformed Weibull hazard, interpolating between stretched-exponential and power-law tail behavior. It reduces to the classical Weibull for $q \to 1$ and to $q$-exponential for $k=1$.

The fit of the q-Weibull distribution is clearly better for cases (a)-(e) of Fig.~\ref{fig:CCDFs}, which is also documented by the Kolmogorov-Smirnov statistics~\cite{HollanderM-2013}, whose values are given in the descriptions of the corresponding panels. Characteristically, the parameter $q$ in these fits is clearly greater than 1, signaling fatter tails of these empirical distribution than the conventional Weibull can fit.
For case (f), the $q$-Weibull distribution gives essentially the same fit as the Weibull distribution $(q=1)$. Fitting with the three-parameter q-Weibull distribution for empirical distributions with jumps, as in cases (g)-(i), is not advisable because the fit may fall into local minima and give completely erroneous results, and therefore is not performed in these cases. 


\section{Temporal correlations}


\begin{figure*}[!htb]
    \centering
    \includegraphics[width=1\textwidth]{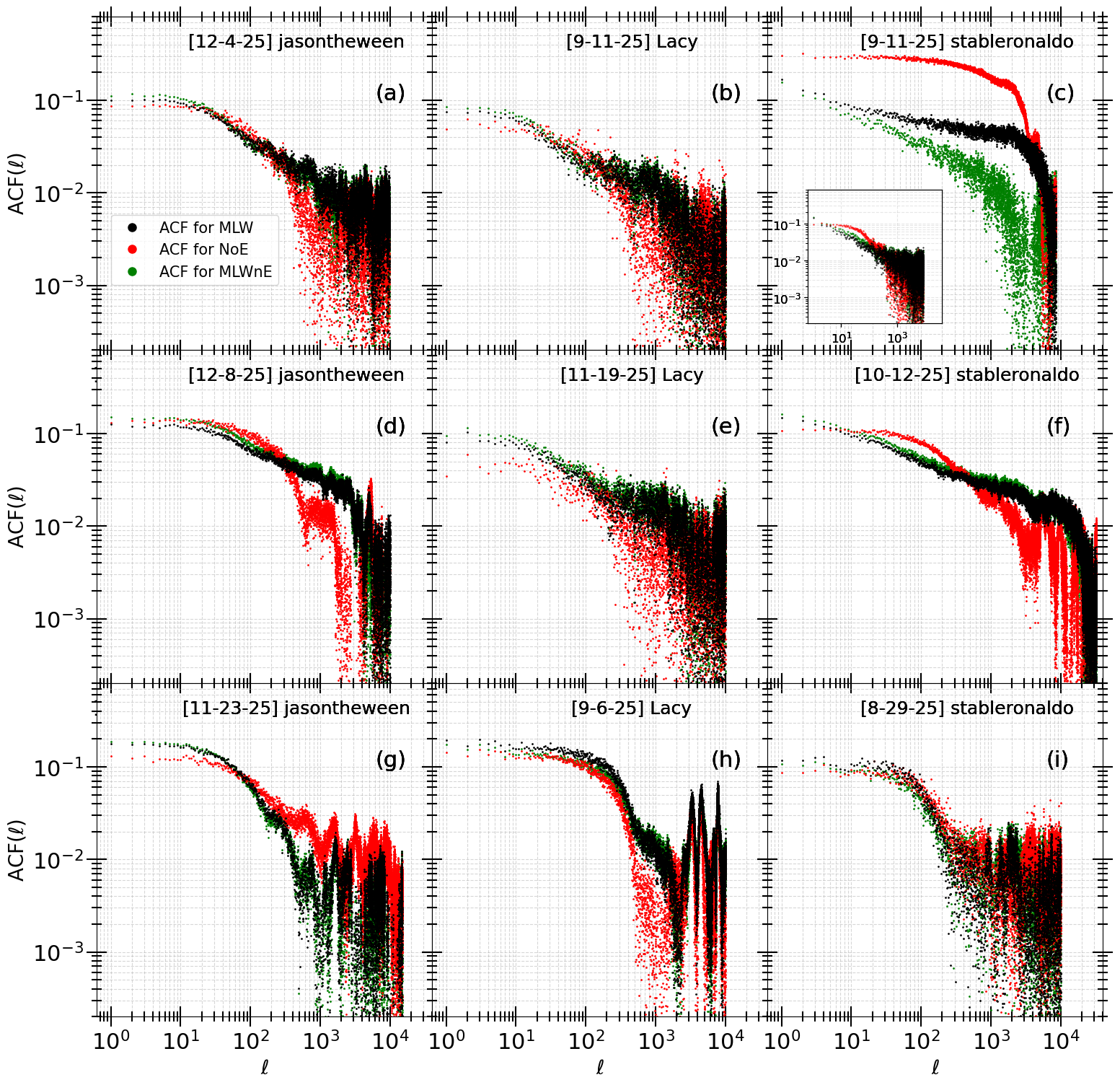}
    \caption{Autocorrelation functions (ACF) for all the nine chats whose message length variability are shown in Fig.~\ref{fig:wideplot}. MLW refers to all the words and emotes included in the messages, MLWnE to words with no emotes included and NoE to only emotes taken into account.}
    \label{fig:ACF}
\end{figure*}

While the distributions of fluctuations provide important information about the range and relative frequency of observed values, they do not reveal how these values are organized in time. To investigate the temporal organization of the series, one must examine temporal correlations, which are commonly quantified by the autocorrelation function (ACF). The ACF measures the statistical dependence between observations separated by a given time lag and thus characterizes the memory and persistence of the underlying dynamics. This information is complementary to that contained in the fluctuation distributions. Indeed, randomly shuffling a time series leaves the distribution of fluctuations unchanged but completely destroys the temporal ordering of the data, causing the autocorrelations at nonzero lags to vanish. Consequently, the ACF encodes dynamical features of the process that are inaccessible from the distribution alone. The ACF used in the present study is defined as 

\begin{equation}
\mathrm{ACF}(\ell) =\frac{\sum_{k=1}^{N-\ell}
(x_k-\bar{x})(x_{k+\ell}-\bar{x}) }{(N-\ell)\sigma^2 },
\end{equation}

\[
\sigma^2
=
\frac{1}{N}
\sum_{k=1}^{N}(x_k-\bar{x})^2,
\]
where $l$ denotes the message lag, $N$ — length of the series, $\bar{x}$ — sample mean and $\sigma^2$ — sample variance.

The ACF results for all 9 series discussed in Figures 1-2, in variants (i) with all words and emotes (MLW), (ii) excluding emotes (MLWnE), and (iii) including only emotes (NoE) are illustrated in Fig.~(\ref{fig:ACF}). 
As can be clearly seen, the correlations expressed by ACF in all these cases are long-range and the positive values of ACF$(l)$ extend even to values of $l$ of the order $10^3 - 10^4$ messages.
The mean rates listed in Table I of Appendix range from about 2.75 to 8.42 messages per second, and the instantaneous rates vary considerably, so there is no fixed conversion from message lag to clock time but on average it can be estimated to assume values of the order of a few minutes. 
However, in most cases the ACFs of different chats differ in important quantitative elements. The most uniform behavior of ACF$(l)$ is for cases (a), (b) and (c) where it falls according to the rather suggestive power law (straight line on the log-log scale) ACF$(l) \sim l^{-0.5}$. However, in case (c), such ACF behavior (shown in the inset of this panel) corresponds to that series starting at $index = 10^4$. In the initial part of this series, the share of emotes is disproportionately large, as can be seen in Figure~\ref{fig:wideplot}. Including also this fragment of the chat into the whole - thus the entire stream - leads to ACF behavior as shown in panel (c). Then, we can see a very large relative amplification of autocorrelation due the emote-related component. Cases (d), (e) and (f) also show a similar long-range character of ACF, although in a less uniform way, especially when it comes to the relative contribution of emotes in building autocorrelation (waveforms marked in black versus red). Finally, the greatest discrepancies in the range of autocorrelation, and especially its homogeneity, are visible in panels (g), (h), and (i). In these cases, ACF has a greater tendency to decline - already at around $l \sim 10^2$ seconds - but also to later oscillatory behavior.

\section{Multiscaling}

The statistical properties of the nine chat series analyzed in the previous sections provide strong motivation for investigating their multifractal characteristics. It has already been demonstrated that the distributions of message-fluctuation lengths exhibit relatively thick tails, indicating the presence of large deviations occurring more frequently than expected for purely random processes. In addition, the temporal organization of these fluctuations reveals persistent long-range autocorrelations extending across broad time scales. Such features are often regarded as signatures of complex dynamics generated by the interplay of multiple temporal mechanisms rather than by a single scaling process. Consequently, it is natural to examine whether the observed chat activity possesses multifractal structure, i.e., whether different fluctuation magnitudes are governed by distinct scaling exponents. 

To address this question, we employ Multifractal Detrended Fluctuation Analysis (MFDFA)~\cite{KantelhardtJ-2002a,OswiecimkaP-2006a}, a well-established method for quantifying multifractal scaling in nonstationary time series. MFDFA is particularly suitable in the present context because it allows one to detect and characterize scale-invariant correlations while effectively removing local trends that could otherwise obscure the underlying dynamics. Moreover, the method provides access to the generalized Hurst exponents and the multifractal spectrum, enabling a comprehensive description of the hierarchy of scaling behaviors present in the data. By examining the dependence of fluctuation functions on both scale and statistical moment order, MFDFA can distinguish between monofractal and genuinely multifractal organization. The application of this approach to the nine chat series therefore offers an opportunity to determine whether the heavy-tailed distributions and long-range temporal correlations identified earlier are accompanied by a richer multifractal structure reflecting the complexity of human communication dynamics. 


\begin{figure*}[!htb]
    \centering
    \includegraphics[width=1\textwidth]{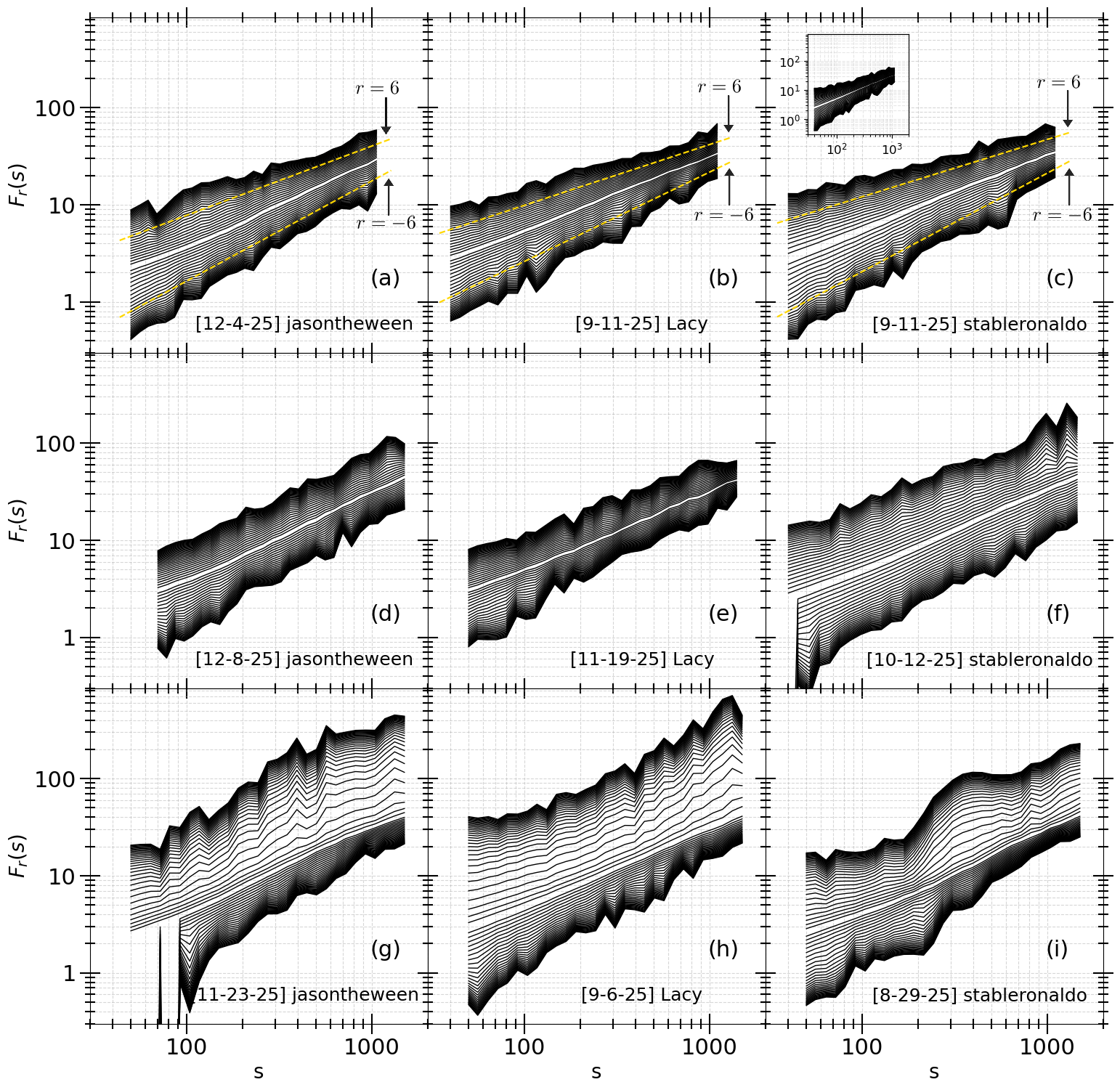}
    \caption{Fluctuation functions for Message Length in Words for nine selected streams. Parameter $r$ lies in range $[-15,15]$ with step $\Delta r = 0.5.$ A second-order polynomial is used here for detrending. Yellow dashed lines indicated by arrows represent fits to $F_{-6}(s)$ and $F_{6}(s).$ Here and subsequently, the entire displayed range of $s$-variation of these functions is used for the calculations.}
    \label{fig:Fr}
\end{figure*}

\subsection{Multifractal Detrended Fluctuation Analysis}

 In formal terms, for a time series $U=\{u\}_{i=1}^T$ of length $T$ whose fractal properties are to be addressed, in the first step of the procedure, one creates a time series profile $\hat{U}=\{\hat{u}_i\}_{i=1}^T$ by integrating $U$
\begin{equation}
\hat{u}_i = \sum_{j=1}^i u_j, \qquad i=1,...,T
\end{equation}
and partitions it into a set of disjoint intervals of length $s$ going from both ends of $\hat{U}$ -- this gives $2M_s$ such intervals total. In each of these intervals $\nu$ a best fitting polynomial trend $P_k^m(\nu)$ of order $m$ is subtracted from $\hat{U}$ (typically $m=2$). This procedure results in a detrended residual signal $X=\{x_i\}_{i=1}^T$
\begin{equation}
x_{s\nu+k} = \hat{u}_{s\nu+k} - P_{\nu}^m(k),
\end{equation}
where $k=1,...,s$ and $\nu=0,...,2M_s-1$. As a next step the variance of $X$ is calculated for each segment:
\begin{equation}
f^2(s,\nu) = \frac{1}{s} \sum_{k=1}^s \left( x_{s\nu+k} - \langle x \rangle_{\nu} \right)^2,
\end{equation}
where $\langle x \rangle_{\nu}$ stands for mean of a segment $\nu$. 


\begin{figure*}[!htb]
    \centering
    \includegraphics[width=1\textwidth]{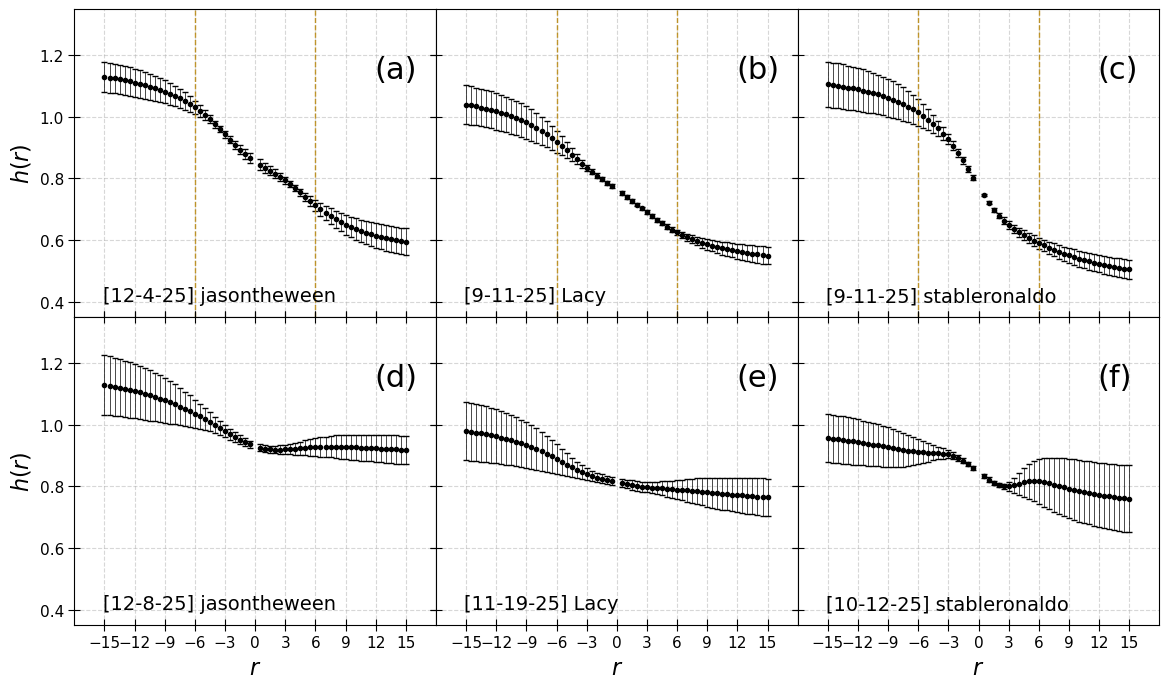}
    \caption{Data points indicate the generalized Hurst exponents $h(r)$ for $r\in [-15,15]$ with step $\Delta r = 0.5$. 95$\%$ confidence intervals of the corresponding fits in terms of Eq.~(\ref{eq:h(r)}) are marked by the shaded areas.}
    \label{fig:Hurstr}
\end{figure*}

Multiscaling is quantified in terms of the family of $r$-dependent fluctuation functions $F_r(s)$ defined by
\begin{eqnarray}
F_r(s) = \biggl\{ \frac{1}{2 M_s} \sum_{\nu=0}^{2M_s-1} \left[ f^2(s,\nu) \right]^{r/2} \biggr\}^{1/r}, \quad r \ne 0,\\
F_r(s) = \exp \bigg\{ \frac{1}{4M_s} \sum_{\nu=0}^{2M_s-1} \ln f^2(s,\nu) \bigg\}, \quad r = 0.
\end{eqnarray}
These functions develop a power-law dependence on scale $s$
\begin{equation}
F_r(s) \sim s^{h(r)},
\label{eq:h(r)}
\end{equation}
for all values of $r$ if $U$ has fractal properties. Otherwise, no such dependence is observed. In practice such a scaling character of $U$ is identified if $F_r(s)$ is well approximated by straight lines on a double logarithmic plot.

For our nine series representing the message length variation of the chats studied here, the fluctuation functions calculated according to the MFDFA procedure presented above are shown in Fig.~\ref{fig:Fr}. As can be clearly seen, they exhibit quite diverse behavior for individual streams. For the first three cases, (a)-(c), the fluctuation functions $F_r(s)$ show very good scaling at first glance (straight-line behavior on a log-log scale), and in addition, it is clearly multifractal, i.e., the slope coefficients of the straight lines significantly depend on $r$, which will be more clearly emphasized later. For the next three streams, (d)-(f), the situation is clearly less impressive. While traces of scaling behavior can be discerned for intervals in $s$, the values of the scaling coefficients will typically depend somewhat on the interval in $s$ considered. Moreover, the dependence of these coefficients on $r$ is also clearly weaker here, signaling a weaker multiscale nature of these dynamics. And no such effects are visible for streams (g)-(i). The fluctuation functions are here much more distorted compared to the previous cases to speak of uniform multiscaling in the sense of the MFDFA formalism.

A function $h(r)$ in Eq.~(\ref{eq:h(r)}) is commonly interpreted as a generalized Hurst exponent of $U$. By mathematical arguments it is a nonincreasing function of $r$~\cite{HentschelHGE-1983}. It is a decreasing function of $r$ if $U$ is multifractal~\cite{HalseyTC-1986} or, in the simplest case, it is constant ($h(r)=H$) which means that $U$ is monofractal~\cite{BarabasiAL-1991}, where $H \equiv h(2)$ is the conventional Hurst exponent of $U$~\cite{HurstHE-1951}. 

For the fluctuation functions of cases (a)-(f) of Fig.~\ref{fig:Fr}, the corresponding $h(r)$ determined by Eq.~(\ref{eq:h(r)}) are shown in Fig.~\ref{fig:Hurstr} for $r \in [-15,15]$. The 95$\%$ confidence interval~\cite{NeymanJ-1937} of the fit by Eq.~(\ref{eq:h(r)}) is explicitly marked. As already mentioned and clearly visible in Fig.~\ref{fig:Fr}, the use of Eq.~(\ref{eq:h(r)}) is not justified for cases (g)-(i) and is therefore not carried out here. In this Figure, we find confirmation of the fact already indicated earlier that the range of variation $\Delta h(r)$ is larger for cases (a)-(c) than for cases (d)-(f). Moreover, the latter cases are burdened with a larger error in determining their values, and in case (d) for positive values of $r$ slightly above 0, $h(r)$ even shows an increasing tendency, which indicates certain numerical problems in applying the MFDFA procedure here. 


\begin{figure*}[!htb]
    \centering
    \includegraphics[width=1\textwidth]{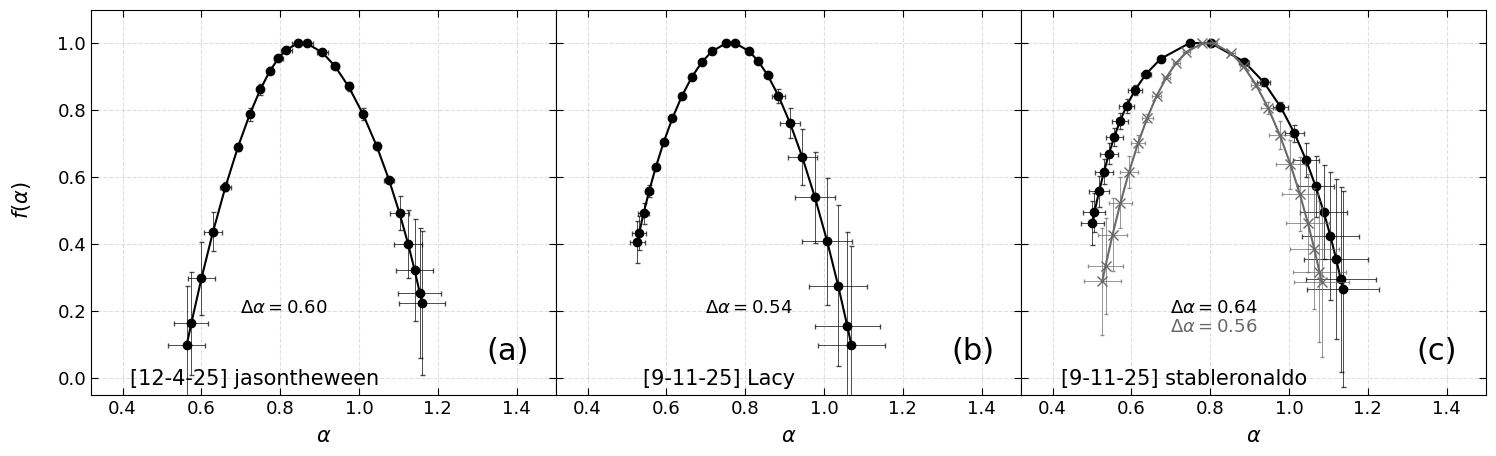}
    \caption{Singularity spectra (black dots) for chats (a)-(c). Moments order parameter $r\in [-6,6]$ with step $\Delta r = 0.5.$ The symbol x in panel (c) indicates the spectrum obtained after deleting the first $10^4$ elements in that series and thresholds of lengths 23 and 25. Error bars indicated are obtained from the confidence intervals of Fig.~\ref{fig:Hurstr}}
    \label{fig:spectra}
\end{figure*}


\begin{figure*}[!htb]
    \centering
    \includegraphics[width=1\textwidth]{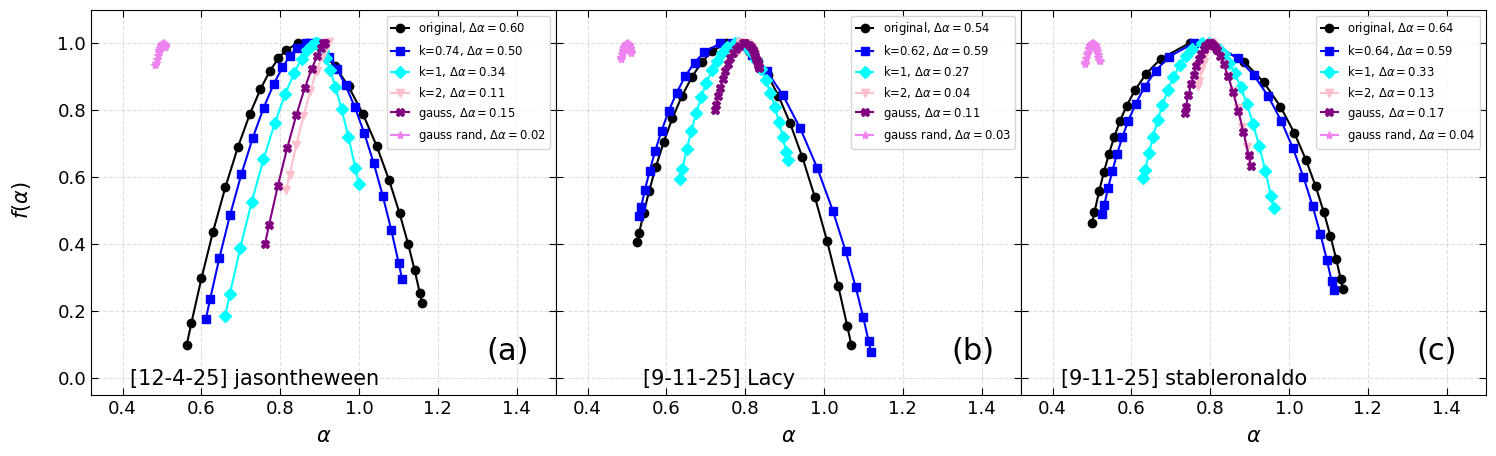}
        \caption{Surrogate singularity spectra for the three cases discussed in Fig.~\ref{fig:spectra} obtained from series constructed in such a way that the temporal autocorrelations of the original series are preserved but the fluctuation distributions are replaced by the procedure proposed in ref.~\cite{KluszczynskiR-2025}. Two types of distributions were used: (i) Weibull for k=1, k=2, and k, which within this distribution gives the best fit to the corresponding empirical data from Fig.~\ref{fig:CCDFs}, and (ii) a half-Gaussian with the maximum located at zero fluctuation values. For this distribution, the multifractal spectrum is also shown when the time series are randomly reshuffled and thus the time correlations are destroyed. As can be seen, the singularity spectrum then becomes like that of white noise.}
    \label{fig:surrogates}
\end{figure*}

In general the functional dependence of $h(r)$ encodes basic information about the multiscale organization of a series and, in particular, about admissible range of the moment order $r$ for which the analysis remains reliable. In the case of leptokurtic probability density functions (PDFs), $ \vert r \vert $ cannot be increased without limit because sufficiently high-order moments may cease to exist. Although finite-size effects may mask such divergences and create the misleading appearance that all moments are finite, the valid range of $r$ can still be assessed. This range can be estimated from the form of the PDF associated with the data. In the light of the results of Section III, the problem of moment divergence in principle may concern the data analyzed here (the $q$-Weibull fit indicates the power-law asymptoticity of the distributions) but only for very high moments of the order of 10. In practice, the range of moments $r$ can be inferred directly from the behavior of $h(r)$. When $h(r)$ displays the characteristic variation expected for a multifractal signal while approaching saturation rapidly for both positive and negative $r$, it suggests that the signal amplitudes are effectively bounded. Under these circumstances, the analysis should be confined to the interval of $r$ over which $h(r)$ retains a noticeable dependence on $r$. This, in particular applies to the singularity spectrum $f(\alpha)$, which is defined as the Legendre transform of the scaling exponent $\tau(r)=r h(r)-1$~\cite{HalseyTC-1986a}:
\begin{eqnarray}
f(\alpha) &=& r\bigl[\alpha(r)-h(r)\bigr] + 1, \\
\alpha(r) &=& h(r) + r \frac{d h(r)}{d r}.
\end{eqnarray}
For a multifractal data set, the function $f(\alpha)$ forms a concave curve, typically exhibiting an approximately parabolic shape over a finite interval of its argument, $\Delta\alpha=\alpha_{\max}-\alpha_{\min}$. In contrast, for a monofractal signal the spectrum collapses to a single point located at $\alpha_0 = H$.

In this framework, it is important to note that genuine multifractality requires the underlying data to exhibit long-range power-law correlations. If this condition is not satisfied, the signal is either monofractal or bifractal, the latter corresponding to a two-point spectrum with support $\{\alpha_1=0,\alpha_2=H\}$~\cite{NakaoH-2000a}. Moreover, even in the absence of long-range power-law correlations the finite-size effects may generate an apparently continuous singularity spectrum, and therefore particular caution is required when interpreting results for which $\Delta\alpha > 0$~\cite{DrozdzS-2009a,KwapienJ-2023a}. 

Another element to keep in mind in the above relationships is that for large values of $ \vert r \vert $, moments are determined by extremely large (for positive $r$) or extremely small (for negative $r$) events in the time series. They can even be influenced by events that are not present in the current sample but could potentially occur. Such virtual effects manifest themselves, for example, as negative fractal dimensions~\cite{mandelbrot1990,ChhabraA-1991}. To avoid entering this area, by direct inspection we limit our analysis to $r \in [-6, 6]$. In the case of the time series analyzed here, for $r > 6$, one often enters the realm of negative fractal dimensions.

For the first three clearly multiscale cases in Fig.~\ref{fig:Hurstr}, the corresponding singularity spectra (also called multifractal spectra) are shown in Fig.~\ref{fig:spectra}. As is typical for model singularity spectra reflecting the complexity of the multiscale organization of the process, they take the form of an inverted parabola and are wide by $\Delta \alpha$ even slightly above 0.5. 
While the spectra in cases (a) and (c) are roughly symmetric, those in case (b) are clearly right-side asymmetric. This type of asymmetry signals~\cite{DrozdzS-2015} that the smaller fluctuations are arranged in a richer multifractal hierarchy than their larger counterparts. Indeed, a look at the corresponding panel (b) of Fig.~\ref{fig:wideplot} shows that the large fluctuations are organized somewhat differently relative to the smaller ones than in the adjacent panels (a) and (c). The resulting CCDF distribution in Fig.~\ref{fig:CCDFs} (b) also has thicker tails than those in (a) and (c) and this may distort the consistent uniform hierarchy of multiscaling proportions.

\subsection{Disentangling sources of multifractality}
Reliable estimation of multifractal properties with MFDFA requires particular caution, especially when the available records are short. In such cases, commonly used surrogate procedures, including random shuffling of time series containing even tens of thousand observations, may still produce apparent multifractal scaling. This apparent scaling can be caused by residual correlation effects rather than by genuine multifractal organization~ \cite{KwapienJ-2023a}. The problem is further amplified for signals with heavy-tailed fluctuation distributions, where finite-size effects may falsely suggest multifractality unless sufficiently long samples are analyzed~\cite{DrozdzS-2009a}. To address this issue more rigorously, a recent validation procedure~\cite{KluszczynskiR-2025} allows to suppress the influence of heavy tails while keeping the temporal ordering, and hence the correlation structure, intact. This is achieved through a rank-based probability-density transformation that gradually maps the empirical fluctuation distribution toward a Gaussian or some other appropriate form~\cite{KluszczynskiR-2025}. 

This type of comparative analysis for the present data, showing how, while maintaining autocorrelation but reducing the thickness of the tails of the fluctuation distribution affects the width of the multifractal spectrum, was carried out and the result illustrated in Fig.~\ref{fig:surrogates}. For this purpose, the Weibull distribution was used, which with increasing $k$ tends to the Rayleigh distribution and reaches it for $k=2$. It is then related to the Gaussian through the radius of a 2D Gaussian vector, so the asymptotic of its tail for large fluctuations is of the Gaussian type. However, for small fluctuation values, this distribution tends to zero. Therefore, as an additional element of this analysis, a half-Gaussian distribution with a maximum located at zero fluctuation values was used. As clearly is seen in Fig.~\ref{fig:surrogates}, the width of the spectra decreases significantly when the distributions have reduced tail thickness. However, they remain multifractal. This property is completely degraded when the distributions are replaced by Gaussian ones and the temporal correlations are destroyed by random shuffling of the series. Thus, without temporal correlations, there is no multifractality also in the case of the data analyzed here. These results also confirm the validity of the numerical procedure used here.

\section{Summary}

The present study examined whether real-time Twitch chat activity, despite its apparent disorder and fragmented form, can exhibit temporal organization characteristic of complex dynamical systems. Twitch chats provide high-frequency records of collective human behavior in which large numbers of users react simultaneously to shared external stimuli, streamer actions, and one another. In this sense, stream chats may be interpreted not only as sequences of individual messages, but also as emergent collective signals shaped by attention, emotion, imitation, and social coordination. From the perspective of social physics, digital communication platforms provide particularly rich settings for investigating collective human behavior using concepts and methods originating in statistical and complex-systems physics \cite{KwapienJ-2012a,JusupM-2022}.

To investigate this possibility, selected high-activity Twitch chats were transformed into time series by measuring the length of each consecutive message in words, with platform emotes treated as individual tokens. This representation captures both linguistic and symbolic components of communication while preserving the temporal order of messages. The analysis focused on nine chat sessions associated with three streamers: jasontheween, Lacy, and stableronaldo. These sessions were selected because they displayed diverse forms of temporal variability and correlation structure.

The empirical message-length series revealed strongly heterogeneous and bursty behavior. Their fluctuation distributions were heavy-tailed, indicating that large deviations occur more frequently than would be expected for simple random or light-tailed processes. Several of these distributions were better described by q-Weibull-type forms than by conventional Weibull or q-exponential fits alone, suggesting the coexistence of stretched-exponential behavior in the central range and heavier tails at larger fluctuation amplitudes. This points to a complex organization of message lengths rather than a single characteristic scale. 

Temporal correlation analysis further showed that the observed variability is not merely a consequence of independent message generation. Autocorrelation functions remained positive over broad ranges of time lags, in some cases extending to time scales of thousands of seconds. This persistence indicates that the stream chats retain memory over substantial periods. The comparison between full message-length series, text-only series, and emote-only series also demonstrated that emotes can play an important role in shaping the temporal structure of chat activity, sometimes strongly enhancing correlation effects.

The main part of the analysis applied Multifractal Detrended Fluctuation Analysis (MFDFA) to assess whether the chat time series display multiscale organization. The results show that distinct, unambiguous multifractality occurs in such series but is not common. In several cases, especially among the first group of analyzed chats, the fluctuation functions displayed clear scaling behavior and the generalized Hurst exponents varied substantially with moment order. The corresponding singularity spectra were broad and well developed, providing evidence for genuine uniform multifractal structure. This, however, is a rather rare property and applies to at most a few percent of chats as estimated by direct inspection of several hundred chats selected by random sampling from a set of approximately $10^5$ chats collected for the analysis presented here. Of course, such a sample is not intended to estimate the population-level frequency of multifractality in all Twitch chats. The results identify possible dynamical regimes rather than a universal property of Twitch communication.

Additional tests involving transformations that reduce the influence of extreme values indicate weakening of the multifractal spectra but not their full elimination. At the same time random shuffling, which destroys temporal order, entirely removes the multifractal structures. This confirms that long-range temporal organization is essential for the emergence of multifractality in these chat signals and it therefore is genuine~\cite{KwapienJ-2023a}. 

Other chats showed weaker scaling, approximate monofractal behavior, or no well exposed single scaling regime. Thus, Twitch chat activity should not be treated as intrinsically multifractal in all cases; rather, multifractality emerges only under particular dynamical conditions. Of course, it is also possible that in such more 'erratic' cases we are dealing with a superposition, or perhaps even simply a mixture, of different multifractal components governed by different cascade generators. However, there is currently no practical procedure that would allow us to disentangle such complex dynamics into individual components.

Overall, the present findings support a cautious interpretation of some Twitch chats as emergent collective signals displaying features structurally reminiscent of stream-of-consciousness narrative. The term stream of consciousness is not used here in its strict narratological or psychological sense, which normally refers to the literary representation of the continuously unfolding inner experience of a single individual. Nor is it intended to imply the existence of a unified collective mind, shared subjectivity, or common intentional agency among chat participants. Rather, the analogy concerns the observable organization of the communication stream: a rapid and only weakly prearranged succession of fragmented utterances, immediate emotional reactions, emotes, repetitions, associative topic shifts, and recurrent motifs, all generated by many heterogeneous users responding both to a common audiovisual stimulus and to one another. Although each message originates from an individual participant, their temporally ordered superposition can produce a distributed, externally stimulated, and endogenously amplified flow in which reactions trigger further reactions, local cascades coexist with slower attention cycles, and apparently disconnected contributions become coupled through imitation, repetition, shared context, and collective affect. In this at present deliberately restricted sense, the expression collective stream of consciousness serves as a structural and dynamical analogy rather than as a literal literary classification. Importantly, the observation of multifractality does not by itself establish semantic, cognitive, or phenomenological equivalence between Twitch chat and literary stream-of-consciousness writing.
However, some closer relationships of this type cannot be ruled out, especially since, as far as has been identified so far~\cite{DrozdzS-2016a,StaniszT-2024a,DolinaM-2025}, multifractal sentence length variability is only found in works belonging to such a style.

The present analysis is based on the temporal variability of message length and therefore concerns the multiscale architecture of the communication process, not the meanings, intentions, or subjective mental states represented by individual messages. What the multifractal results demonstrate is that, in a small subset of chats, the seemingly disordered sequence of contributions possesses a persistent and hierarchically organized temporal structure compatible with such an analogy. This places these chats within a broader class of complex human-generated signals, including certain literary, financial, physiological, and mobility time series, in which local irregularity coexists with organization across multiple scales. A potentially relevant indication is provided by Fig.~\ref{fig:num_of_users}: in the three chats exhibiting the clearest multifractality, the number of active users, following an initial transient, tends to stabilize at a roughly comparable level of about 500, whereas no analogous stabilization is observed in the chats lacking well-developed multifractal scaling. This observation suggests that a sufficiently large but relatively stable population of interacting participants may facilitate the emergence of coherent multiscale organization. It should, however, be regarded as a hypothesis rather than a demonstrated causal mechanism, and its verification will require a substantially broader statistical analysis combining temporal, semantic, interaction-network, and stream-event information.

\section*{Data Availability Statement}

The data that support the findings of this study are openly available in RepOD at
https://doi.org/10.18150/YJGITL, reference~\cite{DataSet}.

\newpage

\section{Appendixes}
\label{Appendix}
\newpage

\begin{longtable*}{@{}p{4cm}|@{ \hspace{1pt} }p{1.22cm}|@{ \hspace{1pt} }p{1.22cm}|@{ \hspace{1pt} }p{1.22cm}|@{ \hspace{1pt} }p{1.22cm}|@{ \hspace{1pt} }p{1.22cm}|@{ \hspace{1pt} }p{1.22cm}|@{ \hspace{1pt} }p{1.22cm}|@{ \hspace{1pt} }p{1.22cm}|@{ \hspace{1pt} }p{1.22cm}|@{}}
\caption{Description of the quantitative characteristics of chats} \label{tab:chat_stats} \\
Statistic &  jason \newline 12-4-25 & jason \newline 12-8-25 &  jason \newline 11-23-25 & Lacy \newline 9-11-25 & Lacy \newline 11-19-25 & Lacy \newline 9-6-25 &  stable \newline 9-11-25   & stable \newline 10-12-25 & stable \newline 8-29-25 \\
\hline
\endfirsthead
\caption[]{Description of the quantitative characteristics of chats} \\
\endhead
\multicolumn{9}{r}{Continued on next page} \\
\endfoot
\endlastfoot
Total number of messages & 107170 & 102396 & 154364 & 81848 & 40713 & 146803 & 73610 & 346878 & 39296 \\
Length in hours & 5.46 & 3.91 & 8.37 & 7.40 & 4.11 & 5.59 & 4.74 & 17.99 & 1.30  \\
Number of chat users & 8046 & 9041 & 9928 & 8491 & 4820 & 16678 & 8665 & 24926 & 5106 \\
\hline
Emotes Ratio & 0.31 & 0.29 & 0.28 & 0.16 & 0.12 & 0.24 & 0.31 & 0.26 & 0.27 \\
  Emote Frequency Rate & 57.08 & 56.55 & 53.93 & 37.43 & 27.79 & 46.93 & 53.68 & 52.01 & 52.22 \\
 Pure Emote Frequency Rate & 44.11 & 41.73 & 41.65 & 26.89 & 18.98 & 36.26 & 43.21 & 38.38 & 42.78 \\
Unique Chatter Volume & 24.56 & 38.54 & 19.77 & 19.12 & 19.55 & 49.73 & 30.47 & 23.09 & 65.46 \\
\hline
mean MLW & 2.06 & 2.26 & 2.19 & 2.64 & 2.63 & 2.37 & 2.51 & 2.46 & 2.20 \\
mean NoE & 0.64 & 0.65 & 0.61 & 0.43 & 0.32 & 0.56 & 0.77 & 0.64 & 0.60 \\
std MLW & 2.09 & 2.34 & 2.58 & 2.85 & 2.60 & 3.90 & 3.21 & 2.95 & 2.69 \\
mean NoM & 5.45 & 7.28 & 5.12 & 3.07 & 2.75 & 7.30 & 4.31 & 5.36 & 8.42 \\
std NoM &  4.36 & 4.57 & 4.02 & 2.84 & 2.45 & 4.28 & 4.21 & 4.08 & 4.31 \\
\hline
\end{longtable*}

\begin{longtable*}{@{}p{3cm}|@{ \hspace{1pt} }p{14.35cm}}
Term/Abbreviation &  Description  \\
\hline
\endfirsthead
\endhead
\endfoot
\endlastfoot
 MLW - Message Length in Words & Sequence of numbers $MLW_i$, $i=1,...,N$, where $N$ is the total number of messages in chat, and $MLW_i$ is the number of words used to write $i$-th message. For example if $j$-th message is "This game is so much fun!", then $MLW_j = 6.$  \\
 \hline
 NoE - Number of Emotes & Sequence of numbers $NoE_i,$ $i=1,...,N$, where $N$ is the total number of messages in chat, and $NoE_i$ is the number of emotes in $i$-th message. For example if $j$-th message is "This game sucks!!!", then $NoE_j = 0,$ and if $(j+1)$-th message is "Let's go!!! CurseLit PogChamp", then $NoE_{j+1} = 2.$ \\
 \hline
 NoM - Number of Messages & Sequence of numbers $NoM_t,$ $t=1,...,T$, where $T$ is the total length of the stream measured in seconds and $NoM_t$ is the number of messages sent by chat users in $t$-th second. For example if $NoM_{200} = 3,$ then in $t=200$th second of the stream users sent in chat $3$ messages. \\
  \hline
 Total number of messages & Total number of messages sent in chat during the stream.  \\
  \hline
  Length in hours  & Length of the stream in hours. \\
  \hline
    Emotes Ratio & Represents the proportion of emotes relative to the total number of words in chat, calculated as \newline $ \text{Emotes Ratio} = \frac{\text{total number of emotes in chat}}{\text{total number of words in chat}}.$ \\
  \hline
 Emote Frequency Rate &  Measures the percentage of total chat messages that contain at least one emote, serving as an indicator of how frequently users supplement or replace text with visual icons ($\text{Emote Frequency Rate} = \frac{\text{number of messages with at least one emote}}{\text{number of all messages}} \cdot 100\%$, or equivalently, it's a percentage of messages for which $NoE_i>0$). \\
  \hline
 Pure Emote Frequency Rate & Quantifies the percentage of the total message volume that contains exclusively emotes, serving as a metric for the frequency of purely symbolic discourse within the chat ($\text{Pure Emote Frequency Rate} = \frac{\text{number of messages with all words being emotes}}{\text{number of all messages} } \cdot 100\% $, or equivalently, it's a percentage of messages for which $NoE_i= MLW_i$). \\
  \hline
 Unique Chatter Volume & Mean number of distinct contributors identified within each one-minute interval of the stream, representing the typical scale of active audience engagement in chat (Unique Chatter Volume $= \frac{\text{number of all (unique) chat users}}{\text{length of the stream in minutes}}$ ). \\
  \hline
    mean MLW &  Mean Message Length in Words - denotes mean number of words used to write each message. \\
  \hline
   mean NoE &  Mean Number of Emotes - mean number of emotes used in every message in chat.  \\
  \hline
   std MLW &  Standard deviation of MLW. \\
  \hline
   mean NoM & Mean number of messages sent by users in each second of the stream; our findings show consistently high activity levels, with average number of messages spanning from 2.75 to 8.42 messages per second. \\
  \hline

\end{longtable*}

\begin{longtable*}{@{}p{3cm}|@{ \hspace{1pt} }p{14.35cm}}
Shortcut &  Stream title  \\
\hline
\endfirsthead
\endhead
\endfoot
\endlastfoot
\hline
 [12-4-25] jasontheween &  JASON X YONNA TWITCH RECAP WASSUP \\  \hline
 [12-8-25] jasontheween &  STREAMER AWARDS RECAP MAYBACH MONDAY FT. TARA YUMMY \\  \hline
 [11-23-25] jasontheween &  LIVE CAMPAIGNING FOR STREAMER AWARDS VOTE JASONTHEWEEN FOR STREAMER OF THE YEAR JUST CHATTING STREAMER DUO  \\  \hline
  [9-11-25] Lacy &  HAWAII STREAM WITH DARLA HELICOPTER RIDE LUAU, GETTING DRUNK \\  \hline
  [11-19-25] Lacy &  RV SLEEPOVER w\_ @Marlon OTW -\_ OKLAHOMA VOTE LACY AND MARLON FOR STREAM DUO OF THE YEAR YOUNGLA 15\% OFF CODE\_ LACY \\  \hline
[9-6-25] Lacy &  FNCS GLOBAL CHAMPIONSHIP WATCHPARTY WITH RON 24 HOUR STREAM SUBSCRIBE FOR NO ADS 35\% OFF SUBS \\  \hline
  \hline
 [9-11-25] stableronaldo &  Explaining everything on why I stopped streaming [twitter\_insta stableronaldo] !snap !com !sub \\
  \hline
  [10-12-25] stableronaldo &   BIGGESST CAR MEET OF THE YEAR BROKE MY SUB RECORD FAZE SUBATHON 2 DAY 12 BALD AT 100k [twitter\_insta @stableronaldo] \\
  \hline
 [8-29-25] stableronaldo &  LOST IN TEXAS  [twitter\_insta stableronaldo] !snap !com !sub \\

\end{longtable*}

\nocite{*}
\bibliography{references}

\end{document}